\documentclass[acmsmall,screen]{acmart}

\usepackage{soul}
\usepackage{multirow}
\usepackage{caption}
\usepackage{subcaption}
\usepackage{xspace}
\usepackage{amsmath}
\usepackage{algorithm,algorithmic}
\usepackage{fancybox}
\usepackage{hyperref}
\usepackage{graphics}
\usepackage{tabularx}

\usepackage{longtable}

\usepackage{rotating}
\usepackage{lscape}
\usepackage{todonotes}

\newcommand{\rebuttal}[1]{\textcolor{blue}{#1}}

\newcommand{\ourtitle}{An Empirical Study on Code Review Activity Prediction and Its Impact in Practice}

\usepackage{xcolor,colortbl}
\newcommand{\bestcell}{\cellcolor{blue!25}}

\newcommand{\RQone}{Which text embeddings of code diffs perform best for review activity prediction?}
\newcommand{\RQtwo}{Which review process features perform best for review activity prediction?}
\newcommand{\RQthree}{How does the combination of review process features with text embeddings perform for review activity prediction?}

\title{\ourtitle}

\setcopyright{rightsretained}
\acmDOI{10.1145/3660806}
\acmYear{2024}
\copyrightyear{2024}
\acmSubmissionID{fse24main-p790-p}
\acmJournal{PACMSE}
\acmVolume{1}
\acmNumber{FSE}
\acmArticle{99}
\acmMonth{7}
\received{2023-09-28}
\received[accepted]{2024-04-16}

\begin{document}


\author{Doriane Olewicki}
\affiliation{%
  \institution{Queen's University}
  \city{Kingston}
  \country{Canada}}
\email{doriane.olewicki@queensu.ca}

\author{Sarra Habchi}
\affiliation{%
  \institution{Ubisoft}
  \city{Montreal}
  \country{Canada}
}
\email{sarra.habchi@ubisoft.com}

\author{Bram Adams}
\affiliation{%
  \institution{Queen's University}
  \city{Kingston}
  \country{Canada}}
\email{bram.adams@queensu.ca}

\begin{abstract}
	
	During code reviews, an essential step in software quality assurance, reviewers have the difficult task of understanding and evaluating code changes to validate their quality and prevent introducing faults to the codebase. This is a tedious process where the effort needed is highly dependent on the code submitted, as well as the author’s and the reviewer’s experience, leading to median wait times for review feedback of 15-64 hours. Through an initial user study carried with 29 experts, we  found that re-ordering the files changed by a patch within the review environment has potential to improve review quality, as more comments are written (+23\%), and participants' file-level hot-spot precision and recall increases to 53\% (+13\%) and 28\% (+8\%), respectively, compared to the alphanumeric ordering. Hence, this paper aims to help code reviewers by predicting which files in a submitted patch need to be (1) commented, (2) revised, or (3) are hot-spots (commented or revised). To predict these tasks, we evaluate two different types of text embeddings (i.e., Bag-of-Words and Large Language Models encoding) and review process features (i.e., code size-based and history-based features). Our empirical study on three open-source and two industrial datasets shows that combining the code embedding and review process features leads to better results than the state-of-the-art approach. For all tasks, F1-scores (median of 40-62\%) are significantly better than the state-of-the-art (from +1 to +9\%). 
	
\end{abstract}



\maketitle

\section{Introduction}
In software development, code reviewing is an essential quality assurance step during code integration~\cite{rigby2012contemporary,fagan2002history}. 
Developers submit their code for review as a patch. 
Then, reviewers either validate the code change, which is then forwarded to integration, or they provide feedback. 
The review process can go through several iterations, where the reviewer comments on the submission to have the developer revise it and submit a new version, referred to as a revision.
Eventually, if the developer addresses all comments, the patch is accepted, otherwise it is rejected.

Given the essential role of code reviews in modern development~\cite{sadowski2018modern}, it is crucial to have a high code review velocity and effectiveness. 
Yet, Hong et al. report that developers waited a median of 15-64 hours for reviewers' feedback in three open source projects (i.e., Openstack Nova, Openstack Ironic and Qt Base)~\cite{hong2022should}.  
Those numbers were confirmed by our industrial partner company, which has similar waiting times.
Such long waiting times delay the integration of code and thus the projects' development and release time. 
Furthermore, other factors can cause review processes to take a long time, including the reviewers involved (e.g., availability and experience), the submitted code (e.g., patch size and complexity, developers’ experience) and the number of review iterations needed~\cite{kononenko2016code}.
Reviewers are usually developers on the project, who have to split their work time between their own development tasks and reviews.
Furthermore, they might have to review many patches, themselves composed of many files and changes. 

In order to improve review effectiveness, Hong et al.~\cite{hong2022should} automate the identification of the parts of patches that either need revisions or comments using a Bag-of-Words (BoW) embedding and two review process features (i.e., number of lines added and removed). 
As BoW embeddings do not represent the order of words such as Large Language Model (LLM) embeddings preferred in recent NLP-based tasks~\cite{hou2023large} do, we suspect that the former do not capture enough context. 
With approaches such as risk/bug prediction using software process features~\cite{nayrolles2018clever,giger2012method,kamei2012large}, it is also unknown how that type of embedding performs in the context of review activities. 

Furthermore, 
while all patch 
revisions are assumed to map to a corresponding review comment, 
revisions could also be triggered by a review comment on a different file or patch, or through communication external to the reviewing platform (e.g., in-person), yielding an incomplete ground truth. Hence, to be comprehensive, automation should not only target problematic parts of a patch requiring either a comment or a revision, but the union of these, which 
we refer to 
as \textbf{hot-spots}.

Prediction of review activity needs has potential to assist reviewers through multiple use cases, such as adding warnings in the code for reviewers to focus on~\cite{henley2018cfar} or re-ordering files~\cite{huang2018salient,petrovic2018state}. In this paper, we focused on the latter. In a wizard-of-oz user study with 29 professional developers, we find that, inspired by Fregnan et al.'s suggestion to ``display first those files that are more critical''~\cite{fregnan2022first}, hot-spot based file re-ordering correlates with improved review comment quality. In particular, 
more review comments were provided (+23\%), and those comments were better targeted towards known hot-spots (+13\% precision and +8\% recall) compared to the alphanumeric ordering. Even though reviewers would still be required to look at the rest of a patch, automated identification of hot-spots would attract the reviewer's attention earlier to the most critical parts
.

Hence, in order to improve the file-level prediction of three review tasks (i.e., need for comment, need for review, is an hot-spot) in terms of prediction performance and of the computation cost of 
embeddings and model training, our paper studies the impact of text embeddings and of review process feature-based embeddings. 
We focus on file-level prediction due to its natural fit with the use case of 
prioritizing files in a submitted patch. 
Our empirical evaluation on three large open-source datasets and two closed-source datasets addresses the following 
research questions: 

\textbf{RQ1: \RQone}
Although none of the specific text embedding stands out as the best, we found that applying any (or both) of our text embedding modifications, i.e., including removed lines as an additional context embedding or using a Large Language Model embedding instead of the Bag-of-Word approach, improved the performances in practice. 
In practice, our approaches report higher AUC in 13 out of 15 cases (median of 68-78\%, improvement up to +5\%) and higher F1-score in 10 out of 15 cases (median of 37-57\%, improvement up to +4\%) than the baseline.

\textbf{RQ2: \RQtwo}
Closed-source datasets show no significant differences between the best text embedding and review process feature embedding (median F1-score 37-56\%, with an improvement in practice of +2 to +8\%), thus matching performance while reducing the embedding dimensionality. 
On the other hand, for open-source datasets, feature-based embeddings perform similarly to better than the best text embedding in 5 out of 9 cases (median F1-score of 50-58\%, improvement up to +4\%) and lower in the other 4 cases (median F1-score 36-56\%, lower by -2 to -15\%).
This shows that text embeddings have value as they capture relevant code/text information, not represented in the feature-based embeddings.
We also find that the size-based features are the most important features for all datasets, as they bring information about the changed file (e.g., size, number of changes) but also context about the full patch size. 
The history-based features, though less important, also improve performance.

\textbf{RQ3: \RQthree}
Compared to the state-of-the-art, our proposal of combining text embedding with review process features improved significantly the F1-score (median of 40-62\%) for all datasets and all tasks, improving the results by up to +9\%.

Finally, through an additional empirical analysis we confirm the user study's promise of hot-spot based file re-ordering, as our model predictions push +9-50\% more hot-spots forward into the first half/quarter of file patches shown to reviewers compared to alphanumerical order.

This paper's contributions are: 
\begin{itemize}
    \item A user study evaluating hot-spot-based file re-ordering;
    \item An empirical study of review activity prediction per-project and at file-level on 5 large datasets (including 2 closed-sources);
    \item 3 types of review activity labels: revised, commented, and hot-spot (either of the prior);
    \item A comparison of text embeddings from a Large Language Model and Bag-of-Words embedding, as well as the inclusion of only added lines or also including removed lines; 
    \item Analysis of lightweight review process features;
    \item A replication package with the open-source dataset~\cite{replic}.
\end{itemize}

\section{Background and Related Work}
For years, the software engineering domain has proposed approaches to help developers in tasks such as defect prediction~\cite{borg2019szz, nayrolles2018clever, olewicki2024costs}, code recommendation~\cite{takuya2011spontaneous}, 
and others~\cite{kim2013should, xia2017effective,da2016framework,anvik2006should, olewicki2022towards}. 
In the last 10 years, the focus has moved towards helping reviewers as well, regarding subjects such as reviewer recommendation~\cite{thongtanunam2015should,rahman2016correct, thongtanunam2014improving}, code change ordering~\cite{baum2017optimal}, code diff quality~\cite{li2022automating,hong2022should}, review comment recommendation~\cite{hong2022commentfinder,li2022automating, tufano2022using, siow2020core,gupta2018intelligent, tufano2021towards}, review comment evaluation~\cite{bosu2015characteristics, rahman2017predicting, efstathiou2018code} and code refinement~\cite{li2022automating, tufano2021towards, tufano2022using}.
In this paper, we focus on the subject of code diff quality, by identifying code submissions needing review activity (i.e., revisions and/or comments).
Bosu et al. reported that 34.50\% of code review comments from five major Microsoft projects that they studied are not ``useful'', i.e., did not lead to code modification, thus arguing the need for code review evaluation~\cite{bosu2015characteristics}.
However, we consider that any review activity has a high value.
Submission of a revision indicates that the code needed to be modified due to an issue (e.g., bug or efficiency) or did not follow some code convention of the team. 
Review comments have also value as they might indicate unclarity in the code (e.g., the nomenclature is not clear), the need for changes in another part of the submission, or the need for a future code submission treated separately.
As both revisions and comment have value, we also want to predict "hot-spots" as changes that requires any of those review activities.


Li et al. explored code diff quality in order to identify parts of the code change that need review comments~\cite{li2022automating}. Their approach uses a Text-To-Text Transfer Transformer (T5) that takes as input the code change and returns a probability that the code change needs a comment. 
Their T5 is pre-trained on a dataset composed of on code diffs. 
Hong et al. proposed the ``RevSpot" approach to predict problematic lines, i.e., lines that will be commented on by the reviewer or revised~\cite{hong2022should}. Their approach uses input based on code vocabulary and they train classification models such as Random Forest and Naive Bayes.
This simpler training approach has the advantage of being less costly, whereas fine-tuning a T5 model is known to be computationally expensive. 
The classification-based approach reports similar results on the "commented" prediction task, indicating that the computational effort of fine-tuning an LLM might not be relevant for this task.

We also explore in this paper the use of Large Language Model (LLM) encoders for text embedding.
LLM are models trained to solve Natural Language tasks. 
They distinguish themselves from Language Models (LM) due to the size large number of parameters if the LLM models~\cite{hou2023large}.
The need for large models comes from the observation that the more parameters are added, the better the LM model performs~\cite{wei2022emergent}. 
LLMs are used in three main types of architectures: encoder-only, encoder-decoder and decoder-only.
Encoder-only architectures take text as input and generate an embedding, i.e., an abstracted representation of the code in the form of a vector capturing the text's semantics~\cite{devlin2018bert}.
Encoder-decoder architectures take text as input, then generate an intermediate embedding state that is itself fed as input to the decoder~\cite{vaswani2017attention,han2021transformer}. 
The intermediate state is used as a way to represent the connection between various inputs and outputs.
This type of architecture is usually used for translation tasks or question-answering.
Finally, the decoder-only architectures use a decoder to generate text, usually in a sequential prediction paradigm. 
This type of model has been vastly popularized over the last few years with GPT models~\cite{openai2023gpt4} and ChatGPT~\cite{chatgpt}.

LLMs have been used in the context of Software Engineering for multiple tasks, such as code synthesis/generation~\cite{austin2021program, zan2023large}, review automation~\cite{tufano2021towards, tufano2022using, li2022automating}, and code fuzzing~\cite{deng2022fuzzing}.
In this paper, we focus on the encoder-only type of task, as we use the LLM as a text embedding approach.
The advantage of this is that we can use a pre-trained LLM model as a zero-shot encoder, where we do not fine-tune the LLM.
The embedding is then used as input to a classifier, easier to train, thus giving a more cost-effective and lower computation-time training approach.
Previous studies have found "simpler" approaches such as classification models to be similar, and sometimes even better, than Deep Learning (DL) approaches in the context of software engineering tasks in terms of both performance~\cite{fu2017easy,hellendoorn2017deep} and speed~\cite{majumder2018500}, which motivates our use of classifiers on top of a zero-shot LLM embeddings instead of fine-tuning a LLM encoder-decoder model.


Other past works regarding bug prediction~\cite{giger2012method}, risk prediction~\cite{shihab2012industrial, nayrolles2018clever} and quality assurance~\cite{kamei2012large} have shown good performance by using software process features. 
Those features range from size-based features (e.g., number of lines added, removed), contextual features (e.g., size of the patch), and history-based features (e.g., number of times a file was changed, number of participants on a file), to experience-based features (e.g., experience of developers/reviewers).
However, review process prediction tasks have not considered exploring those types of features yet, apart from Hong et al. considering only the two following features: number of lines added and the number of lines removed~\cite{hong2022should}.
Exploring review process features approaches is thus one of the goals of this paper.


\section{Motivational Use Case}
\label{sec:motivation}

By default, most review tools list the changed files of a patch in alphanumeric order, even though previous work argues that this might be suboptimal, especially for long code submissions~\cite{baum2016need,barnett2015helping}. Yet, it is common, especially in closed-source datasets, to have many file changes in a patch, 
making the review process more difficult and time consuming.
Case in point, across the five datasets studied in section 5, we observed that 38-68\% of patches are composed of more than 2 files, while in the two closed-source datasets, 25-30\% of patches have more than 5 files.

Baum et al. established principles for code change ordering to help reviewers become more effective~\cite {baum2017optimal}, 
while Li et al.~\cite{li2022automating} and Hong et al.~\cite{hong2022should} predict submission (patch) parts that need either review comments or a revision, yet neither of them explore or discuss the use of code diff quality as a way to prioritize the code change order. 
Fregnan et al.'s user study indicated that the earlier a file is shown in a patch, the more likely it is to be commented on
~\cite{fregnan2022first}, hence suggesting to prioritize files that need review activity to the beginning of a patch.

In this motivation section, we empirically evaluate the impact of changed file reordering on the review process, through a user study that assumes the availability of models able to perfectly predict review hot-spots (i.e., any file requiring a comment or revision). If successful, this motivates the need for improving current review activity prediction models.

\subsection{User study set-up}

We carried out a wizard-of-oz user study~\cite{dahlback1993wizard} to compare hot-spot-based ordering of a patch's changed files to the typical alphanumeric ordering~\footnote{The study was approved by an ethical board. The certificate number is not disclosed to respect double-blind.}. The wizard-of-oz design means that we used knowledge about the actual hot-spots in the oracle 
(obtained similarly as in Section~\ref{sec:label}), which is useful to evaluate the promise of hot-spot prioritization for file reordering in optimal conditions compared to the alphanumeric ordering used by most review tools~\cite{fregnan2022first}. 
The user study was held on the premises of our industrial partner, and involved 29 developers with an average experience at the company of 6 years, knowledgeable both in developing and reviewing for the studied closed-source project, 
which is actively developed 
by more than 800 developers, with 3-4k patches per month.

We structure the study as an A/B study according to which each participant is asked to review two patches, one patch ordered alphanumerically ("Alph"), and one patch ordered based on hot-spots ("Hot"). 
We paired up participants such that each pair had to review the same two patches (P1,P2), but with different file ordering to control for learning bias. For example, the first participant of the pair would see P1-Alph followed by P2-Hot, while the second would see P1-Hot then P2-Alph. To increase the variety of patches, each pair of participants would work on different sets of patches.

In terms of participant recruitment in the project under study, we selected amongst the top 50\% most active reviewers in the last 6 months 
who also have been relatively active in the last month (i.e., amongst top 40-50\% reviewers). The latter choice aimed 
to find participants experienced in performing reviews, but not overloaded, in order to improve the chances of them participating. We then created pairs of participants that have worked on the same subsystems. For each pair, we also identified a fallback option, i.e., a third participant that could be invited if one of the original participants in the pair is unavailable or unresponsive. The sets were constructed in such a way that all three participants had worked on common subsystems of the studied project. 

Each set was then associated with two patches (from the participants' common subsystems) that none of the participants had submitted or reviewed before. 
We selected only patches of exactly five files, similar to Fregnan et al.~\cite{fregnan2022first}. This choice is motivated by the need for patches large enough to observe the impact of ordering files while keeping the review effort manageable to attract more participants. 
The selected patches were not older than 2 months prior to the study, had received reviews, and needed at least one later revision or involved at least one thread of review comments.

The user study was performed via a mock online review platform. 
The main page explained to participants that they should not access any other platform to help them perform the review, and provided 
access to the 2 patches to review. The actual review pages had a user interface mimicking the review tool used at the company, but customized to allow us to change the file ordering. We also incorporated back-end logic to monitor the start and end time of reviews and to record when each changed file was visible on the screen during the review.

The user study happened over the span of 15 work days, during which we contacted 16 pairs (i.e., 32 participants for 16 patches) via our industrial partner's internal chat application. 11 spare participants from the invited pairs' corresponding set had to be contacted. Out of the 43 contacted participants, 29 performed the study, yielding 55 reviews in total (28 with alphanumeric ordering and 27 for hot-spot-based ordering, thus, not all participants did both reviews). 20 patches were reviewed with both ordering setups. 
The recruitment message mentioned that the participation would take around 20 minutes (although most people took longer) and provided a link to our custom review platform.
While the participants were aware of the study being about review processes, they were not explicitly told that we explored the impact of file re-ordering.
Participants were not offered any compensation for participation, but they participated during their work hours.


\subsection{Results}

\textbf{Table~\ref{tab:userstudy_time} shows that the median duration of individual reviews was 11.4 minutes for both orderings, without significant difference in distribution} (Mann-Whitney test: p-val= .39).
Analyzing the review duration came with some challenges, as variations in duration can come from many factors (e.g., participants' review habits, distraction by other tasks). 
4 reviews took more than an hour, yet manual checking of the recorded review events (i.e., opening/closing of files and insertion of review comments) revealed gaps of more than 40min of inactivity for all 4 data points.
We thus ignored those data points for the duration analysis.
When aggregating the duration per patch, 8 patches out of 17 required a longer duration using hot-spot-based ordering compared to 9/17 patches taking longer for alphanumeric. 

\textbf{The order in which reviewers opened changed files tends to follow the provided order.}
Table~\ref{tab:userstudy_time} shows the median/average ranking position of the five (ordered) files of each patch in terms of when a file is opened for the first time. For instance, if in at least half of the patches for a particular ordering file F1 is opened first, its median ranking position would be 1. 
We can see that both the median and average ranking increase with the file number, except for file F5 (hot-spot based). The latter exception seems to be due to some reviewers starting to opening files from the end, and other reviewers opening a C++ file's corresponding header file before opening other files. The average ranking values show that the order can still vary a lot, depending on the reviewers' experience with the given files, the types of files and their review habits.


\textbf{Across patches, we observe an average of 2.9 comments for hot-spot-based compared to 2.3 for alphanumeric (+23\%) order}.
Table~\ref{tab:userstudy_com} shows that the total amount of submitted comments on the hot-spot-based ordering is 77 comments, 12 comments more than the alphanumeric ordering, even though one additional alphanumeric patch was reviewed. Among both sets of comments, 83\% of the comments seem to be actionable, while the rest comprise comprehension questions.

\textbf{Hot-spot-based ordering gathers 80\% of review comments on the first 3 files, compared to only 58\% for alphanumeric ordering}.
This is somehow expected, as 
the hot-spots historically observed in our datasets by definition are a good indication of where comments are needed.

\textbf{The participants' review comments on hot-spot ordered patches show an increase in precision (+13\%) and recall (+8\%) for finding hot-spots}.
By analyzing if the comments 
(1) are equivalent to ground truth comments or (2) pinpoint the location of a later revision, 
we found that 40 comments (51\%) of the hot-spot-based ordering identified actual hot-spots compared to only 27 (41\%) of the alphanumeric order. On average, 1.5 comments per patch found a hot-spot in the case of hot-spot-based ordering compared to 1 per patch for the baseline (+54\%).

It should be noted that two participants indicated to us that they felt ``something wrong'' with the (hot-spot based) file ordering.
In 3 out of the 27 reviews with hot-spot-based ordering that included C++ code, a participant initially missed the header file (extension ``.h''), wrote a comment about it missing, then canceled the comment after seeing the missing file later in the ordering. 
This shows us that certain files forcibly should be shown together even when using hot-spot-based ordering.

\begin{table}
    \caption{User study's patch run-through analysis: depending on the ordering approach, we report the number of patches reviewed, the median (Med.) and Average (Ave.) duration of the review, and the median and average ranking position of the files' initial opening during the reviews.}
        \label{tab:userstudy_time}
    \centering

    \begin{tabularx}{\textwidth}{c|l|ll|XXXXX|XXXXX}
        \hline
        \multirow{3}{*}{\textbf{Ordering}} & \multicolumn{3}{c|}{\textbf{\#Patches}} & \multicolumn{10}{c}{\textbf{Ranking position of files' first opening}}\\\cline{2-14}
        &\multirow{2}{*}{\textbf{Tot.}} &\multicolumn{2}{c|}{\textbf{Duration}}&\multicolumn{5}{c|}{\textbf{Median}}&\multicolumn{5}{c}{\textbf{Average}}\\
        \cline{3-14}
         &&\textbf{Med.}&\textbf{Ave.}&  F1 & F2 & F3 & F4 & F5&  F1 & F2 & F3 & F4 & F5 \\\hline
         \textbf{Hot-spot-based}&27&11.4m&13.6m&1&2.5&3&4&3&2.26&2.96&3.14&3.26&3.30\\\hline
         \textbf{Alpha}&28&11.4m&10.4m&1&2&3&3.5&4&2.25&2.92&2.93&3.21&3.42\\
    \hline
    \end{tabularx}
\end{table}
\begin{table}
    \caption{User study's review comment analysis: depending on the ordering approach, we report the total number of comments, the comments' position (files F1 to F5 or general Gen), the number of comments that found hot-spots, and the number of hot-spot files that were commented on.}
        \label{tab:userstudy_com}
    \centering    
    \begin{tabularx}{\textwidth}{c|l|XXXXXl|c|Xcc}
        \hline
        \multirow{3}{*}{\textbf{Ordering}} & \multicolumn{8}{c|}{\textbf{\#Comments}}& \multicolumn{3}{c}{\textbf{\#Hot-spot files commented}}\\\cline{2-12}
        & \multirow{2}{*}{\textbf{Tot.}}&    \multicolumn{6}{c|}{\textbf{Position}}&\textbf{Found}&\multirow{2}{*}{\textbf{Tot.}}&\multirow{2}{*}{\textbf{Precision}}&\multirow{2}{*}{\textbf{Recall}}\\ \cline{3-8}
         &  & F1 & F2 & F3 & F4 & F5 & Gen&\textbf{Hot-spot}&\\\hline
         \textbf{Hot-spot-}& \bestcell& \textbf{36\%} \bestcell & \textbf{32\%} \bestcell & 12\% & 5\%& 6\%& \textbf{5\%}  \bestcell & \textbf{51\%} \bestcell  &  \bestcell  &53\% \bestcell & 28\% \bestcell  \\
         \textbf{based}& \multirow{-2}{*}{\bestcell \textbf{77}}  & (28) & (25) & (10) &(4) & (5) &(5)& (40)&\multirow{-2}{*}{ {23}}\bestcell &(out of 43) & (out of 80)\\ \hline
         \textbf{Alpha-}& \multirow{2}{*}{65}& 27\% &15\%  &\textbf{16\%} \bestcell  &\textbf{23\%} \bestcell &\textbf{15\%} \bestcell  &1\%  & 41\% & &40\% &20\%\\
         \textbf{numeric}&   & (18) &(10)  & (11)& (15) & (10) & (1) & (27) &\multirow{-2}{*}{ {18}} &(out of 44) & (out of 88)\\                                             
    \hline
    \end{tabularx}
\end{table}

\noindent\doublebox{%
    \parbox{.97\textwidth}{%
        \textbf{File ordering 
        has an impact on the review process. While there is no impact on review duration, the hot-spot-based ordering correlates with more comments (+23\%) targeted at parts of the code known to need revising or commenting (+13\% precision and +8\% recall). This implies that it is important to improve prediction models of review hot-spots.
        }
    }%
}


\section{Methodology for Empirical Study of Hot-Spot Prediction}
Based on the findings of our user study that file re-ordering during review activities is helpful, 
we want to explore how to improve the models for the three common review activity prediction tasks (i.e., reviewed, commented, hot-spot). 
In particular, we empirically evaluate 
prediction models using different variations of two kinds of embeddings (i.e., text-based and feature-based) on 5 large datasets, amongst which 2 closed-source datasets.

\subsection{Labeling}
\label{usecase}
\label{sec:label}

During review activities, reviewers comment on a patch, i.e., a code submission.
Those patches involve one to several commits, where the first commit is the initial submission and the follow-up commits are revisions.
Each commit contains changed files, which are composed of code line additions and/or deletions. 
Line modifications are lines that receive both a deletion and addition.
Most diff tools will show the before and after versions of the file.

Our review activity prediction tasks extend existing code diff quality prediction works~\cite{li2022automating, hong2022should} in order to highlight changed files that most likely should be commented on or revised. 
Past approaches have used a binary labeling based on if the changed file was commented on or revised~\cite{li2022automating, hong2022should}. 
Some works have also addressed the question of identifying useful comments, as they consider that commented changes that do not lead to revisions are not useful~\cite{bosu2015characteristics}.
However, we consider such comments to still have value, as they can lead to future code submission or discussions.

Hence, in this paper, we consider similar labels ("commented" and "revised") as prior works, but also define a "hot-spot" label, which indicates whether a code change has any review activity (i.e., comment, revisions, or both).
This is because, while comments are meant to lead to revisions, and revisions to be caused by comments, review comments might indicate the need for revisions elsewhere in the patch or changes in future independent patches.
Furthermore, we observe that 10-17\% of patches across our datasets contain changes that are revised, while never commented, typically because of a review comment made on another file or outside the reviewing platform. 
If considering only patches that were commented on, those cases would thus be missed, 
hence the importance of considering hot-spot prediction. 
The labels' definitions can be found in Table~\ref{tab:labels}.
This labeling is done at a file-level granularity, following Hong et al.'s approach~\cite{hong2022should}.

\begin{table}[t]
    \centering
    \caption{Label descriptions (file-level).}
    \label{tab:labels}
    \begin{tabularx}{\textwidth}{l|X}
    \hline
    \textbf{Label}&\textbf{Description}\\\hline
         Commented& Changed file receiving a review comment on any of its lines.\\\hline
         Revised& Changed file for which at least one of its lines will be revised in a future patch.
         \\\hline
         Hot-spot& Changed file that is either commented, revised, or both. \\
         \hline        
    \end{tabularx}
    
\end{table}

\subsection{Datasets}

\begin{table}[t]
    \centering
    \caption{The revision activity in the studied datasets.}
    \label{tab:datasets}
    \begin{tabularx}{\textwidth}{l|c|rXXX|rXXX}
        \hline
         \multirow{3}{*}{\textbf{Project}} & \multirow{3}{*}{\textbf{Period}} & \multicolumn{4}{c|}{\textbf{\#Patches}} &\multicolumn{4}{c}{\textbf{\#Changed files}} \\
         && 
         \multicolumn{1}{c}{\textbf{All}} & \textbf{Com-} & \textbf{Revi-} & \textbf{Hot-} & 
         \multicolumn{1}{c}{\textbf{All}} & \textbf{Com-} & \textbf{Revi-} & \textbf{Hot-}\\
         &&  
         & \textbf{mented} & \textbf{sed} & \textbf{spot} & 
         & \textbf{mented} & \textbf{sed} & \textbf{spot} \\
         \hline
         Base& 2 years & 5,390   & 32\% & 31\% & 42\% & 18,950 & 13\% & 18\% &23\%\\
         Ironic& 3 years & 1,050 & 45\% & 38\% & 53\% & 4,193 & 16\% & 26\% &31\%\\
         Nova& 3 years & 3,678 & 54\% & 42\% & 63\% & 11,997 & 26\% & 30\% &43\%\\
         Closed1& 5 years & 34,511 & 43\% & 49\% & 63\% & 270,396 & 10\% & 25\% &28\%\\
         Closed2& 6 years & 40,236 & 45\% & 49\% & 64\% & 390,075 & 9\% & 24\% &27\%\\
         \hline
    \end{tabularx}
    
\end{table}

In this paper, we use three open source datasets and two closed source datasets provided by our industrial partner (different from the dataset used in the user study). 
The open source datasets we consider are those of Hong et al.~\cite{hong2022should}, i.e., Qt Base (Base), Openstack Ironic (Ironic), Openstack Nova (Nova), 
which were used for commented and revisioned prediction tasks~\cite{hong2022should, tufano2021towards}. 
We then used Gerrit's API to gather extra information about the changed files, i.e., author id, reviewer ids, and full file diff (since their replication package only has the added lines), needed for our embeddings.

The closed-source datasets were provided by our industrial partner, and contain reviews of two projects developed over the course of 5-6 years that used a traditional review process. 
Though we are not able to share these datasets, we are interested in showing those results as the datasets are much larger, go over longer periods and contain a denser review activity, as shown in Table~\ref{tab:datasets}.

More information on the datasets can be found in Table~\ref{tab:datasets}, which only considers the initial commit of each patch and its related changed files.
In the context of review activity prediction, we make predictions on the initial commits only. 
Although the follow-up commits/revisions are necessary in our dataset for labeling, they are not part of the training or test datasets.
The frequency of the three labels at the file and patch levels shows that open-source projects tend to have smaller patches in terms of changed files as they have on average 3-4 times more changed files than patches. 
Closed-source datasets, on the other hand, have 9-10 times more changed files than patches, thus larger patches.
Also, as the patches are larger, closed-source datasets also report slightly lower review comment ratio (9-10\% compared to 13-26\% for open-source). 
For all datasets, we observe that the granularity of file changes compared to patches makes the labels more imbalanced.


\subsection{Data splitting}
\label{sec:datasplit}

 \begin{figure}
    \centering
    \includegraphics[width=.8\textwidth]{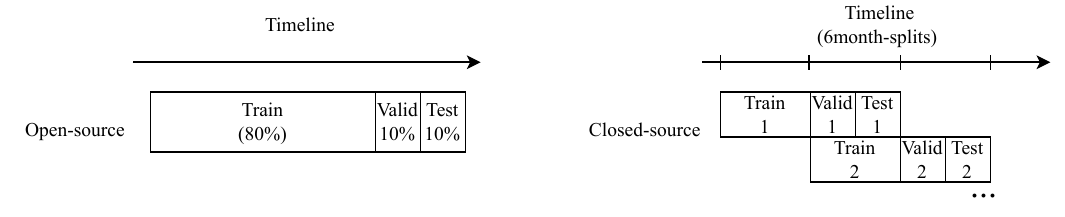}
    \caption{Data split for open- and closed-source datasets.}
    \label{fig:datasplit}
\end{figure}
 
As shown in Table ~\ref{tab:datasets}, the open- and closed-source datasets have different sizes, as the open-source datasets have 4k to 18k changed files, whereas the closed-source datasets have 270k to 390k. 
Since we are considering, among others, Bag-of-Words embedding approaches, training on the full closed-source datasets is not feasible, as the vocabulary of the corpora is too large to maintain.
For this reason, we need different data-splitting approaches for both sources of data. 
However, both splitting approaches respect the data's timeline: changed files present in the test have to be posterior in time to all the changed files in the training set in order not to contaminate the training set, same as for the validation set. Both splitting approaches are shown in Figure~\ref{fig:datasplit}.

In the case of the open-source datasets, we split each dataset along its timeline into a training set validation set and test set of size 80\%-10\%-10\% respectively, similarly to Hong et al.'s approach~\cite{hong2022should}.

In the case of the closed-source datasets, we split each dataset into 6-month periods. Closed1 has 10 such periods and Closed2 has 12. 
From preliminary analysis, we observed that a 6-month period size still allows us to apply the Bag-of-Words approach while keeping a reasonable vocabulary size. 
Then, we apply a sliding window where we train on a period and validating on the first half of the following period and test the second half.
Each training yields an individual model, giving 9 trained models for the Closed1 dataset and 11 for Closed2.

\subsection{Embeddings}

\begin{figure}
    \centering
    \includegraphics[width=.8\textwidth]{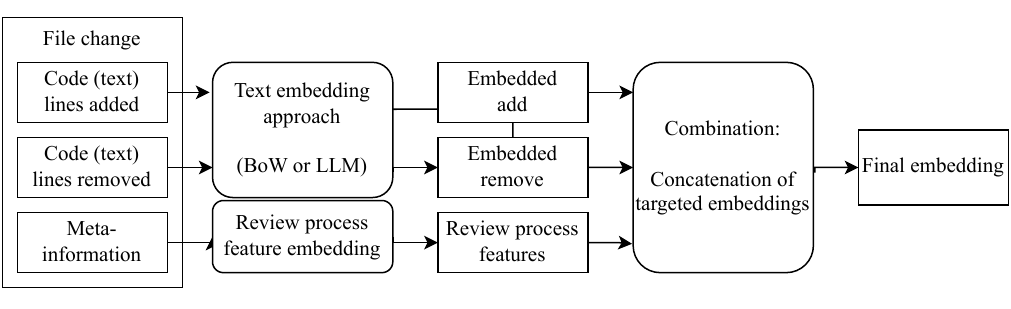}
    \caption{Embedding approach. The file change is composed of lines added, lines removed and meta-information about the submission (e.g., timestamp, author and reviewer ids, patch id). The combination will concatenate the targeted embeddings, i.e., the text embedding, features, or both. The final embedding for each file is 1D.}
    \label{fig:embedding}
\end{figure}

In the context of this paper, we consider two types of text embeddings, i.e., Bag-of-Words (BoW) and Large-Language Model (LLM) encoding, and several review process feature sets.

Figure~\ref{fig:embedding} illustrates the embedding process.
For each changed file, we generate all the embeddings, both text and feature-based. 
At the combination stage, we choose which embeddings to consider, text, feature-based or both.
In the case of text embeddings, we have the choice between only considering added lines or also removed lines. 
In the latter case, added and removed lines are still embedded separately and concatenated at the combination stage. 
For RQ1, we only consider the text embeddings ("embedded add" and "embedded remove"), and we don't take into account any review process features. 
RQ2 considers only the review process features, but the specific features selected depend on the chosen feature setup.
For RQ3, both text embeddings and review process features are considered together and combined using a concatenation approach. 
In each case, the final embedding of a file is a one-dimensional vector.

\subsubsection{Bag-of-Words}

The BoW approach represents each input as a vector of word frequency where each column represents a word in the corpus vocabulary and the value in the vector at that column indicates the number of occurrences of the word in the input.
Considering code changes as input, we have to clean the input in this context, since the source code of changes have many symbols that are not text.
We follow the state-of-the-art's approach~\cite{hong2022should} for code pre-processing, which only considers alphanumeric terms. 
Numbers are also replaced by a token "\texttt{<NUMBER>}".
Still following this approach, we do not perform any stemming, lowercase or lemmatization transformation as such approaches are valuable for natural languages used by humans but do not make sense in the case of code. 
Regarding the BoW computation itself, we use the \texttt{Countvectorize} function of the Scikit-Learn Python library~\cite{contvect}. 
The \texttt{Countvectorize} function is trained on the training set, then is applied to the test set for each data split.
Words that appear in the test set but not in the training set will thus not be included in the bag of words.
To reduce the vocabulary size, we remove words that appear only in one single changed files in the corpus as well as words that are too common (i.e., appear in more than 50\% of the changed file corpus).

\subsubsection{Large-Language Model encoding} 

To work with LLM models, the Hugging Face community provides many pretrained models available to use as-is~\cite{huggingface}. 
We selected Bloom from Big Science, an LLM trained on many programming languages (e.g., Java, C++, Python), who's direct use is text generation and exploring charateristics of language~\cite{bloom}. 
The advantage of Bloom is that it provides a condensed version of the model that still provides good results for NLP tasks. 
We do use \texttt{big science/bloom-560m} in the context of this paper, which is about 4~GB compared to the normal version of Bloom, which is about 375~GB.
Also, it counts more than 1.4M downloads in the month of August 2023, making it the most popular LLM pre-trained on programming languages at the time of our empirical study.
To obtain the encoding, we used \texttt{SentenceTransformer}, a Hugging Face Python library~\cite{sentencetrans}.
For the encoding step, we do not clean the code, as Bloom is trained on raw code. 

\subsubsection{Review process features}
software tasks such as risk prediction~\cite{nayrolles2018clever, shihab2012industrial} and bug prediction~\cite{giger2012method, kamei2012large} have been shown to use software process features for their training and getting respectable performance.  
Based on their example and inspired by their features, we defined review process features following two main categories: size-related features and historical features.

The count features refer to metrics considering the size of the changed file targeted by the prediction, as well as contextual features regarding the size of the changed files' patch.
Historical features refer to the past activities that happened surrounding the targeted file, but also the experience of the code's author and of the patch's reviewers known at the time of the prediction.
The definitions of our features can be found in Table~\ref{tab:meta_feat}.
Those features are generated on each individual dataset, based on the timestamps of the patch submission.

\begin{table}[t]
    \centering
    \small
    \caption{Description of review process features generated at the file-level granularity. <LAB> refers to the different labels we consider.}
    \label{tab:meta_feat}
    
    \begin{tabularx}{\textwidth}{cll|X}
        \hline
        &&\textbf{Feature name}&\textbf{Description} \\ \hline
        \hline

        \parbox[t]{2mm}{\multirow{12}{*}{\rotatebox[origin=c]{90}{Count}}}
         &1)& c\_add & \# lines added in the file.\\
         &2)& c\_rem & \# lines removed in the file.\\
         &3)& c\_unchanged & \# lines unchanged shown in the file. For Gerrit, at most 10 lines of context are shown above and under changed lines as context. In the tool used by our industrial collaborators, the context is at most 5 lines. \\
         &4)& c\_add\_rev & \# lines added in the patch of the file.\\
         &5)& c\_rem\_rev &\# lines removed in the patch of the file.\\
         &6)& c\_unchanged\_rev &\# lines unchanged in the patch of the file.\\
         &7)& r\_add\_rev & $\frac{c\_add}{c\_add\_rev}$ \\
         &8)& r\_rem\_rev & $\frac{c\_rem}{c\_rem\_rev}$  \\
         &9)& r\_unchanged\_rev & $\frac{c\_unchanged}{c\_unchanged\_rev}$  \\
         \hline 
        
        \parbox[t]{2mm}{\multirow{8}{*}{\rotatebox[origin=c]{90}{File's history}}}
         &10)&file\_exp& \# past and current patches that included changes to this file (min = 1).\\
         &11) &dir\_exp& \# past and current patches that included changes to this file's directory (min = 1).\\
         &12- 14)& r\_<LAB>\_file & File's average <LAB> ratio before the current patch (default: 0, value between 0 and 1). \\
         &15- 17)& r\_<LAB>\_dir & The directory of the file's average <LAB> ratio before the current patch (default: 0, value between 0 and 1). \\
         \hline
         
        \parbox[t]{2mm}{\multirow{11}{*}{\rotatebox[origin=c]{90}{Author's history}}}
         &18)&auth\_exp &\# past and current patches submitted by the author (min = 1).\\
         &19)&authfile\_exp & \# past and current patches in which the current file was submitted by the author(min = 1). \\
         &20)&authdir\_exp & \# past and current patches in which the current file's directory was submitted by the author (min = 1). \\
         &21- 23)& r\_<LAB>\_author & Author's average <LAB> ratio before the current patch (default: 0, value between 0 and 1). \\
         &24- 26)& r\_<LAB>\_authfile & Author's average <LAB> ratio on the current file before the current patch (default: 0, value between 0 and 1). \\
         &27- 29)& r\_<LAB>\_authdir & Author's average <LAB> ratio on the current file's directory before the current patch (default: 0, value between 0 and 1). \\
         \hline
         
        \parbox[t]{2mm}{\multirow{8}{*}{\rotatebox[origin=c]{90}{Reviewers' history}}}
         &30)&max\_reviewers\_exp &Maximum \# past and current patches reviewed by one of the reviewers out of all this patch's reviewers (min = 1).\\
         &31) &mean\_reviewers\_exp &Average \# past and current patches reviewed by one of the reviewers out of all this patch's reviewers (min = 1).\\
         &32- 34)& r\_<LAB>\_max\_ reviewers &  Maximum Reviewers's <LAB> rate before the current patch (default: 0, value between 0 and 1). \\
         &35- 37)& r\_<LAB>\_mean\_ reviewers &  Average Reviewers's <LAB> rate before the current patch (default: 0, value between 0 and 1). \\
         \hline
    \end{tabularx}
    
\end{table}

\subsection{Classification}

Similar to  Hong et al.~\cite{hong2022should}, we considered Random Forest (RF), Naive Bayes (NB), Nearest Neighbors, Logistic Regression, Decision Trees, and off-the-shelf LLMs (\texttt{bloom560m}) to classify the changed files as "revised", "commented" and/or "hot-spots" as three separated tasks.
From preliminary analysis on those classifiers, we narrowed down the list to Random Forest and Naive Bayes, as they performed better in terms of AUC and F1-score for our task, regardless of the used embeddings.
This confirmed Hong et al.'s observation~\cite{hong2022should} that Random Forest gives the best AUC for their task, followed in most cases by Naive Bayes.
During preliminary analysis, we used the off-the-shelf Bloom LLM classifier, yet this did not perform as well as the considered ML models.
As all the datasets are imbalanced considering the changed file granularity, we consider the SMOTE data sampling approach during training~\cite{wang2006classification,chawla2002smote}.
Models are trained on a per-project basis.

In the end, we have thus two different models (RF and NB) with and without the SMOTE sampling approach (models with SMOTE will be referred to as RFs and NBs), making 4 options for model selection. 
Each model is trained on a specific label (i.e., commented, revised or hot-spot), thus needing 12 trained models in total for each dataset-split and each setup of embeddings. 

\subsection{Evaluation}

We empirically evaluate the models for the three tasks using three open-source and two closed-source datasets. 
To do the evaluation, we report AUC, as Hong et al.~\cite{hong2022should}. 
We also report metrics used for imbalanced labeled datasets: F1-score 
(F1 = $\frac{2}{pre^{-1}+rec^{-1}}$) and Geometric Mean (GM = $\sqrt{pre \cdot rec}$), which are computed based on the precision (pre = $\frac{TP}{TP+FP}$) and recall (rec = $\frac{TP}{TP+FN}$). 
Those metrics need a threshold to define when the prediction is considered as true or false. We choose a threshold that maximizes the F1-score on the validation sets, as it captures best imbalance of the datasets~\cite{chase2014thresholding}.

In Machine Learning based approaches, it is best to train and test models several times to verify the robustness to randomness in the approaches or data splits.
For the closed-source datasets, we mentioned in Section~\ref{sec:datasplit} how the dataset is split into several periods and thus runs. 
For the open-source datasets, as they are smaller than the closed-source datasets, only one data split is done, following the process explained in Section~\ref{sec:datasplit}. 
However, each embedding setup and model is run 5 times, with a different seed in each iteration. That seed is used in the model (e.g., random forest takes a random seed) and in the SMOTE sampling when used.

For all the datasets, out of the different models (i.e., RF, RFs, NB, and NBs), we report the model with the highest median F1-score on the validation set, as that is the metric we want to optimize. 
The results shown in the tables are the metrics for the run with median F1-score for a given embedding setup and its best model. 

\subsection{Statistical tests}
\label{sec:stats}

As mentioned in the previous section, we run each setup at least 5 times per dataset for each label and the median results are shown in the different tables.
Hence, we need to perform statistical tests to validate whether or not the differences observed are significantly different.

Since we have a limited number of runs, we decided to group the runs into sub-categories of datasets: open-source datasets (i.e., Base, Ironic and Nova), closed-source datasets (i.e., Closed 1 and 2) and all datasets.
When comparing two setups for a given label, we take all the runs of that setup for all the datasets in the subcategory.
We apply a Wilcoxon paired test~\cite{cuzick1985wilcoxon} with $\alpha$-value = .01.
We calculate the Cliff's Delta effect to evaluate the impact of the significant differences~\cite{macbeth2011cliff}.

As an example, let us consider the comparison of two setups for the one label and the sub-category open-source datasets.
Each dataset in that case has 5 runs of the best model, i.e., runs B1-5 for Base, I1-5 for Ironic and N1-5 for Nova. For each setup, we align the runs into a 
vector yielding one vector by setup of the form $[B1-B5~I1-I5~N1-N5]$, in this case of length 15.
Then, we perform the pairwise test between both vectors where a p-value inferior to .01 is reported as significantly different, in which case we also compute the effect size on the same vectors.






\section{RQ1: \RQone}
\subsection{Motivation}

In their paper, Hong et al. explore the use of the Bag-of-Words (BoW) to embed code changes~\cite{hong2022should}. 
Contrary to previous work using full-code changes to perform prediction~\cite{wattanakriengkrai2020predicting}, they only consider added lines of a code change.
This choice has the advantage of not having removed or unchanged lines contaminating the text embedding, as those lines could appear as noise in the training.
However, the prediction will thus have less context since only the added lines are considered. 

Furthermore, the BoW approach, though easy to implement and easily explainable, presents several disadvantages.
First, this approach does not scale with large datasets as the size of the vocabulary increases exponentially with the corpus size.
Secondly, BoW does not capture the essence of source code, as the order of words does not impact the embedding.
It captures the number of occurrences of words and thus the size of the code as well. 

On the other hand, LLM pre-trained encoders are made to represent input data in a representative way, as to capture the underlying structure and semantics of the code. 
Also, the input encoding is a fixed-size embedding. 
However, the difficulty with these approaches is that models are more difficult to explain based on their feature importance, as the abstraction of the original input into the encoded output is not understandable to humans.


\subsection{Approach}

In RQ1, we explore how text embeddings impact the review activity predictions at file-level. 
Following the findings of Hong et al.~\cite{hong2022should}, we will explore both (1) keeping only the added lines of a code change and (2) consider embedding removed lines in addition to the add lines.
Among the different variations of each text embedding, we will identify the best embedding for a given dataset and label as "Best text" embedding.

Table~\ref{RQ1:res} shows the results of both considered embeddings, BoW and LLM encoding, with the variations considering only added lines (+) or both added lines and removed lines (+/-).
We performed a statistical pairwise test using Wilcoxon Test, comparing "BoW +" embedding to each text embedding for each sub-category of datasets (i.e., open-source, closed-source, or both), as mentioned in Section~\ref{sec:stats}. 


\subsection{Results}

\begin{table}[t]
    \centering
\small
    \caption{Results for RQ1: text embeddings for the code review activity prediction task \rebuttal{at file-level}. The results shown are returned for the best model ("mod") across each variation of text embedding. Blue cells indicate the maximum values for each metric per dataset and per label.}
    \label{RQ1:res}
    \begin{tabularx}{\textwidth}{Xl|XXXXXc|XXXXXc|XXXXXc}
       \hline
        &\multicolumn{1}{c|}{\textbf{Text}}
        &\multicolumn{6}{c|}{\textbf{Commented}}& \multicolumn{6}{c|}{\textbf{Revised}}& \multicolumn{6}{c}{\textbf{Hot-spot}}
        \\
        & \multicolumn{1}{c|}{\textbf{Embed}}
        &\textbf{auc}&\textbf{f1}&\textbf{gm}&\textbf{pre}&\textbf{rec}&\textbf{mod}
        &\textbf{auc}&\textbf{f1}&\textbf{gm}&\textbf{pre}&\textbf{rec}&\textbf{mod}
        &\textbf{auc}&\textbf{f1}&\textbf{gm}&\textbf{pre}&\textbf{rec}&\textbf{mod}
        \\

\hline 
\parbox[t]{2mm}{\multirow{4}{*}{\rotatebox[origin=c]{90}{Base}}}
&BoW +
&67& 38& 63& 27& 63&RF&67& 42& 59& 40& 44&RFs&66& 49& 62& 42& 59&RFs\\ \cline{2-20}
&BoW +/-
&68& 40\bestcell & 64& 30\bestcell & 58&RFs&67& 44& 62& 37& 55\bestcell &RFs&67& 50& 62& 41& 63\bestcell &RFs\\ 
&LLM +
&69\bestcell & 39& 65\bestcell & 28& 69\bestcell &RFs&68\bestcell & 46\bestcell & 62& 42\bestcell & 51&RFs&68& 51& 64\bestcell & 44\bestcell & 60&RFs\\ 
&LLM +/-
&69\bestcell & 40\bestcell & 65\bestcell & 28& 69\bestcell &RFs&68\bestcell & 46\bestcell & 63\bestcell & 39& 52&RFs&69\bestcell & 52\bestcell & 63& 44\bestcell & 61&RFs\\

\hline 
\parbox[t]{2mm}{\multirow{4}{*}{\rotatebox[origin=c]{90}{Ironic}}}
&BoW +
&77& 46\bestcell & 67& 39& 53&RFs&70\bestcell & 53\bestcell & 66\bestcell & 41& 81\bestcell &RFs&71\bestcell & 57& 63& 45& 82\bestcell &RF\\ \cline{2-20}
&BoW +/-
&78\bestcell & 45& 69\bestcell & 36& 62\bestcell &RFs&69& 50& 64& 42\bestcell & 68&RFs&71\bestcell & 57& 64& 47& 76&RF\\ 
&LLM +
&78\bestcell & 45& 66& 39& 52&RFs&67& 52& 63& 38& 75&RFs&70& 58\bestcell & 65& 47& 76&RFs\\ 
&LLM +/-
&78\bestcell & 40& 61& 40\bestcell & 45&RFs&67& 51& 62& 37& 77&RFs&70& 58\bestcell & 66\bestcell & 48\bestcell & 76&RFs\\

\hline 
\parbox[t]{2mm}{\multirow{4}{*}{\rotatebox[origin=c]{90}{Nova}}}
&BoW +
&75& 54& 67& 56& 52&RFs&72& 48& 66& 37& 69&RFs&72& 57\bestcell & 46& 42\bestcell & 91&RFs\\ \cline{2-20}
&BoW +/-
&75& 55\bestcell & 68\bestcell & 58& 53\bestcell &RFs&73& 50& 67& 41\bestcell & 67&RFs&72& 57\bestcell & 48\bestcell & 42\bestcell & 90&RFs\\ 
&LLM +
&78\bestcell & 55\bestcell & 67& 57& 51&RFs&75\bestcell & 50& 68\bestcell & 40& 69&RFs&76\bestcell & 57\bestcell & 35& 40& 98\bestcell &RF\\ 
&LLM +/-
&78\bestcell & 53& 66& 59\bestcell & 48&RFs&75\bestcell & 51\bestcell & 67& 41\bestcell & 73\bestcell &RFs&76\bestcell & 57\bestcell & 37& 41& 97&RFs\\

\hline 
\parbox[t]{2mm}{\multirow{4}{*}{\rotatebox[origin=c]{90}{Closed1}}}
&BoW +
&74& 38\bestcell & 62& 31\bestcell & 48&RFs&70& 46& 65& 37& 60&RFs&67& 51& 63\bestcell & 41& 67&RFs\\ \cline{2-20}
&BoW +/-
&74& 36& 65\bestcell & 28& 50&RFs&71& 44& 64& 37& 57&RFs&68& 51& 61& 40& 73&RFs\\ 
&LLM +
&75\bestcell & 37& 62& 29& 53&RFs&72\bestcell & 47\bestcell & 66\bestcell & 36& 70\bestcell &RFs&65& 55\bestcell & 55& 42\bestcell & 79\bestcell &RF\\ 
&LLM +/-
&74& 37& 63& 27& 54\bestcell &RFs&72\bestcell & 45& 64& 38\bestcell & 63&RFs&69\bestcell & 53& 63\bestcell & 41& 68&RFs\\

\hline 
\parbox[t]{2mm}{\multirow{4}{*}{\rotatebox[origin=c]{90}{Closed2}}}
&BoW +
&72& 32\bestcell & 60& 26\bestcell & 43&RFs&70& 39& 63& 34\bestcell & 49&RFs&67& 48& 61& 36& 77\bestcell &RFs\\ \cline{2-20}
&BoW +/-
&75& 32\bestcell & 64\bestcell & 23& 50\bestcell &RFs&71& 40& 65\bestcell & 32& 60&RFs&67& 47& 63& 36& 72&RFs\\ 
&LLM +
&72& 32\bestcell & 62& 23& 49&RFs&70& 41\bestcell & 61& 30& 61&RFs&66& 49& 58& 40\bestcell & 69&RFs\\ 
&LLM +/-
&77\bestcell & 32\bestcell & 58& 24& 42&RFs&73\bestcell & 40& 65\bestcell & 28& 66\bestcell &RFs&72\bestcell & 51\bestcell & 65\bestcell & 40\bestcell & 72&RFs\\

\hline

    \end{tabularx}
\end{table}

\textbf{Compared to "BoW +", no specific text embeddings report significant differences in F1-score across all datasets, for any labels (p-value = .012 to .441).}
The only case that returned significant improvement with small effect size is "LLM +/-" compared to "BoW +" for the label "hot-spot" (p-value = .001), with an improvement of up to +4\%. 
Also, we notice a significant improvement with small effect regarding AUC with embedding "BoW~+/-" for all datasets on labels "commented" and "revised" (p-value = .0).

\textbf{However, when comparing the "Best text" embedding with the "BoW~+", the F1-score is found to be  significantly better with small effect size for labels "revised" (p-value = .007) and "hot-spot" (p-value = .007)}. 
Out of those 10 cases (5 datasets $\times$ 2 labels), 8/10 cases show higher F1-score (from +2 to +4\%), 1/10 case show no effect and one got lower results (-1\%).
The label "commented" does not perform significantly better (p-value = .021). It at least matches in 3/5 datasets (up to +2\%) and is lower in two cases (-1\%).
Also, out of the 15 cases, 8/15 cases got higher F1-score results than the baseline for "LLM +/-", 10/15 cases for "LLM +" and 6/15 for "BoW +/-".



We thus conclude that in general, none of the text embedding stands out in all cases, although in practice, choosing the right variations improves F1-score in 10/15 cases. 
The scalability of BoW still stands, as the vocabulary quickly explodes as the corpus increases in size.

\textbf{For BoW, we found that 95-99\% of the feature importance is focused on 1-5\% of top features, compared to 50\% of feature importance focused on 20-40\% of top features for the LLM embedding.}
We apply a feature importance analysis using Python's \texttt{feature\_importances\_}, which measures the percentage of feature importance of each individual feature. 
Comparison of the "BoW +" and "LLM +" embeddings shows that due to the very large size of the embedding outputs of BoW, the classifier will only focus on a low number of features, with the majority of stored features not playing an important role in the prediction. More details on feature importance can be found in the replication package~\cite{replic}.


\noindent\doublebox{%
    \parbox{.97\textwidth}{%
        \textbf{RQ1: Compared to Bag-of-Words applied only on added lines, adding the removed lines and/or using the Large Language Model embedding significantly improved the F1-score for labels "revised" (up to +3\%) and "hot-spot" (up to +5\%) at file-level. The label "commented" has no significant improvement using the new text embeddings.
        }
    }%
}

\section{RQ2: \RQtwo}
\subsection{Motivation}

The state-of-the-art approach only considers two basic software process features, i.e., number of lines add and removed, which were able to improve the performance of commented and revised prediction tasks~\cite{hong2022should}.
We thus want to evaluate how review process features in general, such as size-related features, but also features based on the history of the file, the author and the reviewers, perform compared to text embeddings, inspired by other software processes that have used similar features in the past~\cite{shihab2012industrial, kamei2012large}.


\subsection{Approach}

In Table~\ref{tab:RQ2_res} we report the results of different combinations of features:
\begin{itemize}
    \item "Add/Rem": this set of features refers to the features used by Hong et al. in their approach~\cite{hong2022should}, i.e., "c\_add" and "c\_rem" in Table~\ref{tab:meta_feat}.
    \item "Count": described in the "Count" section of Table~\ref{tab:meta_feat}.
    \item "Hist": described in the File's , Author's and Reviewers' history sections of Table~\ref{tab:meta_feat}.
    \item "All feat": all the aforementioned review process features, thus all features in Table~\ref{tab:meta_feat}.
\end{itemize}

In RQ2, we only consider review process features, excluding any text embedding, as this is the topic of RQ3. 
However, we will also report in the table the best results of text embedding from RQ1 as "Best text", per dataset and predicted label in order to see how feature-based performs compared to text embeddings, which (theoretically) have more specific information about the file's content.
Again we will define a best setup "Best feat" as being the local best feature-based embedding for each dataset and label. We also applied statistical tests to compare the "Best text" embedding with the feature-based embedding. 


\begin{table}[t]
    \centering
    \small
    \caption{Results for RQ2: software review process features embeddings on the code review activity prediction tasks at file-level. The results are returned for the best model ("mod") for each setting of text embedding. Blue cells indicate the maximum values for each metric per dataset and per label.}
    \label{tab:RQ2_res}
    
    \begin{tabularx}{\textwidth}{Xl|XXXXXc|XXXXXc|XXXXXc}
    \hline
        &\multicolumn{1}{c|}{\textbf{Text}}
        &\multicolumn{6}{c|}{\textbf{Commented}}& \multicolumn{6}{c|}{\textbf{Revised}}& \multicolumn{6}{c}{\textbf{Hot-spot}}
        \\
        & \multicolumn{1}{c|}{\textbf{Embed}}
        &\textbf{auc}&\textbf{f1}&\textbf{gm}&\textbf{pre}&\textbf{rec}&\textbf{mod}
        &\textbf{auc}&\textbf{f1}&\textbf{gm}&\textbf{pre}&\textbf{rec}&\textbf{mod}
        &\textbf{auc}&\textbf{f1}&\textbf{gm}&\textbf{pre}&\textbf{rec}&\textbf{mod}
        \\

\hline 
\parbox[t]{2mm}{\multirow{5}{*}{\rotatebox[origin=c]{90}{Base}}}
& Best text
&68& 40\bestcell & 64& 30\bestcell & 58&RFs&68& 46& 63& 39& 52&RFs&69& 52& 63& 44& 61&RFs\\ \cline{2-20}
&Add/Rem
&59& 33& 58& 25& 50&RF&65& 44& 61& 34& 66&RF&64& 50& 55& 36& 81&RF\\ 
&Count
&65& 36& 59& 24& 73&RF&65& 42& 60& 38& 49&RF&66& 50& 58& 37& 78&RFs\\ 
&Hist
&62& 33& 39& 20& 93\bestcell &RFs&63& 43& 59& 32& 70\bestcell &RF&60& 47& 25& 31& 97\bestcell &RF\\ 
&All
&73\bestcell & 40\bestcell & 65\bestcell & 28& 74&RFs&74\bestcell & 50\bestcell & 65\bestcell & 49\bestcell & 51&RFs&73\bestcell & 54\bestcell & 66\bestcell & 45\bestcell & 70&RFs\\

\hline 
\parbox[t]{2mm}{\multirow{5}{*}{\rotatebox[origin=c]{90}{Ironic}}}
& Best text
&77\bestcell & 46\bestcell & 67\bestcell & 39\bestcell & 53&RFs&70\bestcell & 53\bestcell & 66\bestcell & 41& 81&RFs&70\bestcell & 58\bestcell & 65\bestcell & 47\bestcell & 76&RFs\\ \cline{2-20}
&Add/Rem
&53& 30& 53& 20& 68&RF&61& 45& 43& 31& 90\bestcell &RFs&62& 51& 37& 36& 92\bestcell &RFs\\ 
&Count
&65& 36& 62& 27& 55&RF&61& 44& 58& 38& 49&RFs&63& 52& 56& 40& 72&RFs\\ 
&Hist
&57& 30& 49& 20& 71\bestcell &RFs&70\bestcell & 53\bestcell & 65& 40& 84&RFs&67& 55& 61& 44& 78&RFs\\ 
&All
&67& 36& 62& 25& 67&RF&68& 51& 65& 43\bestcell & 64&RFs&69& 56& 62& 45& 80&RFs\\

\hline 
\parbox[t]{2mm}{\multirow{5}{*}{\rotatebox[origin=c]{90}{Nova}}}
& Best text
&75\bestcell & 55\bestcell & 68\bestcell & 58\bestcell & 53&RFs&75\bestcell & 51\bestcell & 67\bestcell & 41& 73\bestcell &RFs&76\bestcell & 57& 37& 41& 97&RFs\\ \cline{2-20}
&Add/Rem
&63& 40& 39& 26& 88&RF&65& 45& 63& 36& 64&RFs&67& 56& 34& 40& 95&RFs\\ 
&Count
&68& 47& 62& 34& 78&RFs&72& 48& 65& 35& 73\bestcell &RF&68& 58\bestcell & 54\bestcell & 44& 86&RF\\ 
&Hist
&60& 30& 47& 32& 28&RFs&56& 38& 56& 29& 57&RF&58& 41& 53& 49\bestcell & 36&RFs\\ 
&All
&70& 40& 17& 25& 100\bestcell &RFs&69& 47& 62& 45\bestcell & 45&RFs&67& 55& 10& 38& 100\bestcell &RF\\

\hline 
\parbox[t]{2mm}{\multirow{5}{*}{\rotatebox[origin=c]{90}{Closed1}}}
& Best text
&74& 38\bestcell & 62& 31& 48&RFs&72& 47& 66& 36\bestcell & 70&RFs&65& 55& 55& 42& 79&RF\\ \cline{2-20}
&Add/Rem
&63& 24& 59& 19& 56&RF&67& 39& 60& 29& 64&RF&60& 48& 49& 36& 73&RF\\ 
&Count
&74& 36& 64& 28& 50&RFs&75\bestcell & 47& 68& 35& 64&RF&68& 55& 60& 40& 82&RF\\ 
&Hist
&59& 25& 53& 16& 83\bestcell &RFs&58& 38& 47& 24& 83\bestcell &RF&59& 47& 42& 32& 83\bestcell &RF\\ 
&All
&77\bestcell & 37& 66\bestcell & 32\bestcell & 52&RF&75\bestcell & 48\bestcell & 69\bestcell & 35& 67&RFs&69\bestcell & 56\bestcell & 64\bestcell & 43\bestcell & 78&RFs\\

\hline 
\parbox[t]{2mm}{\multirow{5}{*}{\rotatebox[origin=c]{90}{Closed2}}}
& Best text
&77& 32& 58& 24& 42&RFs&70& 41& 61& 30& 61&RFs&72& 51& 65& 40& 72&RFs\\ \cline{2-20}
&\#Add/Rem
&67& 24& 60& 17& 46&RF&65& 35& 61& 23& 64&RF&59& 45& 45& 29& 84&RFs\\ 
&Count
&79& 34& 67& 26& 52&RF&72& 42& 66& 32& 62&RFs&68& 49& 63& 36& 77&RFs\\ 
&Hist
&59& 18& 52& 11& 60\bestcell &RFs&63& 34& 55& 22& 79\bestcell &RF&61& 45& 45& 32& 87\bestcell &RF\\ 
&All
&81\bestcell & 37\bestcell & 69\bestcell & 29\bestcell & 56&RF&76\bestcell & 45\bestcell & 68\bestcell & 35\bestcell & 64&RFs&74\bestcell & 53\bestcell & 67\bestcell & 41\bestcell & 76&RFs\\ 

\hline

    \end{tabularx}
    
\end{table}

\subsection{Results}


\textbf{For the open-source datasets, the "Best feat" embedding reports no significant difference regarding the F1-score for labels "revised" and "hot-spots" (p-value = .804 and .389), thus managing to perform as well as the (supposedly richer) text embeddings.}
We can note that for 3/6 cases, at least one of the feature-based approaches improves F1-score compared to "Best text" (up to +4\% F1-score).
Out of the 6 cases, "All feat" reports better F1-score than the "Best text" in 2/6 cases and "Count" reports a better F1-score in 1/6 cases.
The F1-scores for the label "commented" are significantly lower with medium effect (p-value = .007).
In 4 out of 9 open-source cases, the "Best text" embedding performs better than the feature-based embeddings (+2 to +12\% F1-score).
This is expected, as the text embedding has more specific information than the high-level information of the feature-based embeddings.

\textbf{For the closed-source datasets, the "Count" embedding is not significantly different from the "best text" embedding for any label with respect to F1 score (p-value = .355 to .966). }
Statistically, the text embedding and the "Count" perform similarly though having access to different types of information, the "Count" approach being less computationally expensive as the embedding size is much smaller.
In practice, we observe that in those 6 cases (2 datasets $\times$ 3 labels), "Count" reports at least as well in F1-score to "best text" in 4/6 cases (up to +2\%) and lower in the last 2 cases (up to -3\%).

\textbf{Similarly, for the closed-source datasets, "All feat" does not report significantly different F1-score for any label compared to "best text" (p-value = .034 to .393).}
It reports significantly different AUC for labels "revised" (p-value = .0) and "hot-spot" (p-value = .003) with large effect in both cases.
We observe that "All feat" gets a better F1-score in 5/6 cases ( +1 to +5\%), -1\% in the last case, and it reports better AUC in all 6 cases (+2 to +6\%). 
Furthermore, "All feat" compared to "Count" reports significantly better F1-score with small effect regarding the label "commented" (p-value = .0) and not significantly different for the labels "revised" (p-value = .014) and "hot-spot" (p-value = .032).
However, "All feat" improves F1-score compared to "Count" in all 6 cases, with +1 to +4\%.
In practice, adding the history-based features does improve F1-score and performs better than "Best text".

Through these results, we observe that in the case of open-source datasets, the review process feature-based approach does not consistently perform as well as the "Best text" embedding, though the F1-score reported by the best feature embeddings per dataset and label improved performance in 4/9 cases (up to +12\%). 
Regarding closed-source datasets, the "All feat" embedding performs as well as the "Best text" embedding, while reducing the computational effort of model training thanks to the reduced dimensionality.
Looking at feature importance per feature category of Table~\ref{tab:meta_feat}, 
we found that all datasets benefit most from the "Add/Rem" and "Count" categories, whose features each represent 2 to 13\% of the overall feature importance.  
This indicates that those features regarding the size of changed files have the most value, including also the contextual features regarding the full patch size. More details on feature importance can be found in the replication package~\cite{replic}.


\noindent\doublebox{%
    \parbox{.97\textwidth}{%
        \textbf{RQ2: In the closed-source projects, considering all review process features performs as good as the best text embeddings at file-level in terms of F1-score, while reducing the computational effort as the embedding is of lower dimensionality. 
        For open-source datasets, the best feature embedding per dataset and label matches the F1-score of text embeddings, although no specific feature-based embedding is consistently as good.
        All datasets benefit the most from size-based features.
        }
    }%
}

\section{RQ3: \RQthree}
\subsection{Motivation}

Hong et al. propose in their paper a review activity prediction approach using a BoW embedding in addition to features measuring the number of lines added and removed to a file~\cite{hong2022commentfinder}.
In RQ1 and RQ2, we explored how text embeddings and review process feature embeddings perform separately on the three tasks and what they capture about the code submission. 
In this question, we combine both types of embeddings to find the best embedding, compared to the state-of-the-art.

\subsection{Approach}

We combine each text embedding with each review process feature embedding presented in the previous RQs.
The results are shown in Table~\ref{tab:RQ3_res}. Each project is presented with three comparisons: the baseline (state-of-the-art~\cite{hong2022should}, i.e., BoW embedding with "Add/Rem" features), "Best text" (from RQ1), and "Best feat" (from RQ2).
We report two main combinations: 
first, "Heavy combo" which refers to our most "heaviest" embedding compared to the baseline, i.e., the one considering the "LLM +/-" text embedding and "All feat" feature-based embedding.
Then, we also report "Best combo", which reports the combination with the best F1-score per dataset and label.
The lines directly following "Best combo" report the specific combination chosen (i.e., the name of the text and review process feature embeddings used).

\subsection{Results}



\textbf{For closed-source datasets, "Heavy combo" improves significantly the F1-score compared to the baseline regarding labels "revised" (p-value = .002 with medium effect) and "hot-spots" (p-value = .004 with large effect).}
Out of those 4 cases (2 datasets $\times$ 2 labels), the improvement on F1-score is +3 to +6\%. 
In the case of the label "commented", "Heavy combo" performs better for Closed2 (+4\%) and lower for Closed1 (-3\%).
In all 6 cases, the AUC is improved (+3 to +8\%).
Furthermore, the "Heavy combo" is also the "Best combo" in the case of the label "revised".

\textbf{For open-source datasets, "Heavy combo" is not significantly different than the baseline regarding F1-score for any label (p-value = .048 to .421).}
In practice, it improves F1-score in 5/9 cases (+1 to +6 \%) and AUC in 7/9 cases (+1 to +5\%), but not significantly.

\textbf{For all datasets, "Best combo" performs +1 to +9\% better regarding F1-score than the baseline, with medium effect for labels "commented" and "revised" (p-value = .0) and large effect for large "hot-spot" (p-value = .0).}
"Best feat" significantly improves the F1-score for labels "revised" (p-value = .0 medium effect) and "hot-spot" (p-value = .004 small effect), with improvement up to +8\% for "revised" and +6\% for "hot-spot" compared to the baseline.

Table~\ref{tab:RQ3_res} shows the different embedding choices for the best combinations of text-embedding and feature embedding regarding F1-score. 
We observe that out of the 15 cases, the best setup chosen involved BoW in 6/15 cases and LLM in the other 9. The text embedding including both added and removed lines was chosen in 9/15 cases and only the added lines in 6/15 cases. 
Overall, the classic text embedding "BoW +" was chosen only twice. 
It should also be noted that in 3/15 cases, the "Best feat" embedding got the best results, which indicates that in those cases, combining features with the text embedding did not even improve performance, making the lower-dimensionality review process feature-only approach the most performant.

Regarding the feature embedding choice for "Best Combo", "Add/Rem", the simpler embedding, was never chosen, "Count" was chosen 3/15 times, "Hist" was chosen 4/15 times, and "All feat" was chosen 8/15, making the latter the most popular feature-embedding.

Though the "Heavy combo" embedding is not unanimous, as it improved F1-score for 10/15 cases, the new review process features and text embeddings variations that we consider are worth exploring as their best combination ("Best combo") improves F1-score in all cases.

\noindent\doublebox{%
    \parbox{.97\textwidth}{%
        \textbf{RQ3: 
        The Best combination of embeddings improves significantly the F1-score (+1 to +9\%) for all datasets and labels at file-level.
        Though combining LLM embeddings and all process features is the most common best combination, there is no consensus on the best combination.
        The heavy combination reports significantly higher F1-score for closed-source datasets regarding the labels "revised" and "hot-spots", and in practice improves the F1-score in 10 out of 15 cases across all datasets. 
        }
    }%
}

\begin{table}[t]
    \centering
    \small
    \caption{Results for RQ3: combining the text embeddings with software meta-features embeddings for the code review activity prediction task at file-level. The results are returned for the best model ("mod") for each setting of text embedding. Blue cells indicate the maximum values for each metric per dataset and per label.}
    \label{tab:RQ3_res}
    
    \begin{tabularx}{\textwidth}{Xl|XXXXXc|XXXXXc|XXXXXc}
        \hline
        &\multicolumn{1}{c|}{\textbf{Text}}
        &\multicolumn{6}{c|}{\textbf{Commented}}& \multicolumn{6}{c|}{\textbf{Revised}}& \multicolumn{6}{c}{\textbf{Hot-spot}}
        \\
        & \multicolumn{1}{c|}{\textbf{Embed}}
        &\textbf{auc}&\textbf{f1}&\textbf{gm}&\textbf{pre}&\textbf{rec}&\textbf{mod}
        &\textbf{auc}&\textbf{f1}&\textbf{gm}&\textbf{pre}&\textbf{rec}&\textbf{mod}
        &\textbf{auc}&\textbf{f1}&\textbf{gm}&\textbf{pre}&\textbf{rec}&\textbf{mod}
        \\

\hline 
\parbox[t]{2mm}{\multirow{6}{*}{\rotatebox[origin=c]{90}{Base}}} 
& Baseline
&68& 39& 62& 29& 52&RFs&67& 42& 59& 40& 42&RFs&66& 50& 62& 41& 65&RFs\\ \cline{2-20}
& Best text
&68& 40& 64& 30\bestcell & 58&RFs&68& 46& 63& 39& 52&RFs&69& 52& 63& 44& 61&RFs\\ 
& Best feat
&73\bestcell & 40& 65\bestcell& 28& 74\bestcell&RFs&74\bestcell & 50\bestcell & 65\bestcell & 49\bestcell & 51&RFs&73\bestcell & 54\bestcell & 66\bestcell & 45\bestcell & 70&RFs\\ \cline{2-20}
&Heavy combo
&71& 40& 64& 28& 66&RFs&70& 47& 64& 40& 59\bestcell &RFs&70& 52& 61& 39& 78\bestcell&RFs\\ 
& Best combo
&73\bestcell & 41\bestcell & 6\bestcell5& 29& 71&RFs&70& 47& 64& 40& 59\bestcell &RFs&71& 53& 62& 40& 76&RFs\\  
&&\multicolumn{6}{c|}{BoW +/- $\&$ All}&\multicolumn{6}{c|}{LLM +/- $\&$ All}&\multicolumn{6}{c}{LLM + $\&$ All}\\ 
\hline

\hline 
\parbox[t]{2mm}{\multirow{6}{*}{\rotatebox[origin=c]{90}{Ironic}}} 
& Baseline
&79\bestcell& 46& 69& 37& 58&RFs&71& 51& 65& 41& 69&RFs&72& 57& 65& 47& 74&RF\\ \cline{2-20}
& Best text
&77& 46& 67& 39\bestcell& 53&RFs&70& 53& 66& 41& 81&RFs&70& 58& 65& 47& 76&RFs\\ 
& Best feat
&65& 36& 62& 27& 55&RF&70& 53& 65& 40& 84\bestcell &RFs&69& 56& 62& 45& 80\bestcell &RFs\\ \cline{2-20}
&Heavy combo
&79\bestcell& 44& 63& 41& 47&RFs&67& 48& 60& 36& 74&RF&73\bestcell& 58& 67& 53\bestcell& 67&RFs\\ 
& Best combo
&78& 47\bestcell & 70\bestcell & 38& 64\bestcell &RFs&75\bestcell & 54\bestcell & 67\bestcell & 51\bestcell & 54&RFs&73\bestcell& 60\bestcell & 69\bestcell & 52& 70&RFs\\  
&&\multicolumn{6}{c|}{LLM + $\&$ Count}&\multicolumn{6}{c|}{BoW +/- $\&$ Hist}&\multicolumn{6}{c}{LLM +/- $\&$ Hist}\\ 
\hline

\hline 
\parbox[t]{2mm}{\multirow{6}{*}{\rotatebox[origin=c]{90}{Nova}}} 
& Baseline
&77& 56& 68& 60\bestcell& 52&RFs&71& 47& 65& 38& 62&RFs&73& 57& 49& 42& 88&RFs\\ \cline{2-20}
& Best text
&75& 55& 68& 58& 53&RFs&75& 51& 67& 41& 73\bestcell&RFs&76\bestcell & 57& 37& 41& 97&RFs\\ 
& Best feat
&68& 47& 62& 34& 78&RFs&72& 48& 65& 35& 73\bestcell&RF&68& 58& 54& 44& 86&RF\\ \cline{2-20}
&Heavy combo
&78& 46& 56& 31& 93\bestcell &RF&76\bestcell & 53\bestcell & 69\bestcell & 46\bestcell& 64&RFs&75& 56& 28& 39& 99\bestcell&RF\\ 
& Best combo
&82\bestcell & 61\bestcell & 73\bestcell & 59& 63&RFs&76\bestcell & 53\bestcell & 69\bestcell & 46\bestcell& 64&RFs&75& 62\bestcell & 69\bestcell & 63\bestcell& 63&RFs\\  
&&\multicolumn{6}{c|}{BoW + $\&$ Hist}&\multicolumn{6}{c|}{LLM +/- $\&$ All}&\multicolumn{6}{c}{BoW +/- $\&$ Hist}\\ 
\hline

\hline 
\parbox[t]{2mm}{\multirow{6}{*}{\rotatebox[origin=c]{90}{Closed1}}} 
& Baseline
&73& 38& 63& 32\bestcell& 49&RFs&71& 45& 63& 37& 66&RFs&67& 50& 61& 39& 74&RFs\\ \cline{2-20}
& Best text
&74& 38& 62& 31& 48&RFs&72& 47& 66& 36& 70&RFs&65& 55& 55& 42& 79&RF\\ 
& Best feat
&77& 37& 66& 32\bestcell& 52&RF&75\bestcell& 48& 69\bestcell& 35& 67&RFs&69& 56& 64\bestcell& 43& 78&RFs\\ \cline{2-20}
&Heavy combo
&78& 35& 66& 26& 56\bestcell &RFs&75\bestcell& 50\bestcell & 68& 40\bestcell & 74\bestcell &RFs&70\bestcell& 53& 64\bestcell& 43& 70&RFs\\ 
& Best combo
&79\bestcell & 41\bestcell & 67\bestcell & 32\bestcell& 52&RFs&75\bestcell& 50\bestcell & 68& 40\bestcell & 74\bestcell &RFs&70\bestcell& 59\bestcell & 64\bestcell& 45\bestcell& 80\bestcell&RFs\\  
&&\multicolumn{6}{c|}{BoW +/- $\&$ Count}&\multicolumn{6}{c|}{LLM +/- $\&$ All}&\multicolumn{6}{c}{LLM + $\&$ All}\\ 
\hline

\hline 
\parbox[t]{2mm}{\multirow{6}{*}{\rotatebox[origin=c]{90}{Closed2}}} 
& Baseline
&73& 33& 59& 27& 38&RFs&70& 41& 63& 37\bestcell & 56&RFs&66& 48& 61& 36& 74&RF\\ \cline{2-20}
& Best text
&77& 32& 58& 24& 42&RFs&70& 41& 61& 30& 61&RFs&72& 51& 65& 40& 72&RFs\\ 
& Best feat
&81\bestcell& 37& 69\bestcell& 29& 56\bestcell&RF&76\bestcell& 45\bestcell & 68\bestcell& 35& 64\bestcell&RFs&74& 53& 67\bestcell& 41& 76&RFs\\ \cline{2-20}
&Heavy combo
&81\bestcell& 37& 67& 29& 55&RFs&76\bestcell& 45\bestcell & 68\bestcell& 35& 64\bestcell&RFs&73& 54\bestcell & 65& 41& 78&RFs\\ 
& Best combo
&81\bestcell& 40\bestcell & 68& 33\bestcell & 54&RFs&76\bestcell& 45\bestcell & 68\bestcell& 35& 64\bestcell&RFs&75\bestcell & 54\bestcell & 67\bestcell& 42\bestcell& 79\bestcell&RFs\\  
&&\multicolumn{6}{c|}{BoW + $\&$ Count}&\multicolumn{6}{c|}{LLM +/- $\&$ All}&\multicolumn{6}{c}{LLM + $\&$ All}\\ 
\hline

    \end{tabularx}
    
\end{table}

 

\section{Discussion}

\begin{table}[t]
    \centering
\small
    \caption{Ordering based on hot-spot prediction. We report the "Alph" (alphabetic) ordering, and "Ours" our ordering, based on hot-spot prediction. "\%patch" reports the percentage of patches with the range of file changes. The blue cells indicate the best results by dataset and by metric.}
    \label{discussion}
    \begin{tabularx}{\textwidth}{l|X|XX||X|XX|XX||X|XX|XX}
        \hline
        \multirow{3}{*}{\textbf{Dataset}}
        &\multicolumn{3}{c||}{\textbf{Size 2 to 4}}
        &\multicolumn{5}{c||}{\textbf{Size 5 to 9}}
        &\multicolumn{5}{c}{\textbf{Size 10 and more}}
        \\\cline{2-14}
       & \%  &\multicolumn{2}{c||}{Rec@50\%}  
       & \%  &\multicolumn{2}{c}{Rec@50\%}  &\multicolumn{2}{c||}{Rec@25\%} 
       & \%  &\multicolumn{2}{c}{Rec@50\%}  &\multicolumn{2}{c}{Rec@25\%} \\
       &patch&Alph&Ours
       &patch&Alph&Ours&Alph&Ours
       &patch&Alph&Ours&Alph&Ours
       \\\hline

Base
& 30&  72& \bestcell 81                        
& 7&  72& \bestcell 81&  55& \bestcell 70      
& 3&  67& \bestcell 78&  60& \bestcell 64      
\\                                             
Ironic                                         
& 50&  82& \bestcell 82                        
& 14&  61& \bestcell 90&  44& \bestcell 94     
& 5&  68& \bestcell 77&  63& \bestcell 74      
\\                                             
Nova                                           
& 48&  69& \bestcell 86                        
& 11&  71& \bestcell 89&  53& \bestcell 69     
& 7&  51& \bestcell 73&  31& \bestcell 62      
\\                                             
Closed1                                        
& 38&  61& \bestcell 78                        
& 18&  55& \bestcell 71&  45& \bestcell 64     
& 12&  52& \bestcell 69&  38& \bestcell 59     
\\                                             
Closed2                                        
& 35&  67& \bestcell 79                        
& 14&  57& \bestcell 73&  48& \bestcell 69     
& 11&  54& \bestcell 71&  42& \bestcell 62     
\\                                             

\hline
    \end{tabularx}
\end{table}

\subsection{Hot-spot detection for file re-ordering}

To evaluate the impact of hot-spot prediction-based file ordering (instead of a wizard-of-oz setting, using the ground truth), we took patches with at least 2 changed files and at least one of the files labeled as hot-spot, then made model predictions using RQ3's heavy combination. We ordered each patch's changed files based on their predicted probability of hot-spotness.
We compute the average Recall@k ($=\frac{TP}{TP +FN}$) across patches, where $TP$ are the true positive cases in the top $k$, and $(TP+FN)$ is the total number of hot-spots (maxed at $k$).
k is chosen depending on the size of the patch ($size$), so Recall@50\% means k = $\frac{size}{2}$ and Recall@25\% means k = $\frac{size}{4}$. 
Table~\ref{discussion} compares the recall of our ordering with the alphanumeric order, i.e., the default ordering used by review tools, on patches of different sizes.
\textbf{Our prediction-based empirical analysis shows that hot-spot-based file ordering performs consistently better than alphanumeric ordering in all datasets and for different patch sizes.} Specifically, Table~\ref{discussion} reports a Recall@50\% of 71-90\% for patches of size 5 to 9, and 69-78\% for patches of size 10 and more. Compared to the alphanumeric order, this is an improvement of respectively +9 to +29\% and +9 to +22\%.
When looking at only the top Recall@25\%, our ordering still performs better than the alphanumeric ordering in all cases, improving the results from +13 to +50\% for size 5 to 9, and +4 to +31\% for patch sizes over 10.

In future work, we consider evaluating the use of this hot-spot prediction for file ordering in a set-up similar to the user study of Section~\ref{sec:motivation}, but with a larger scale and over a longer period.

\subsection{Computation time}

One of our motivations for evaluating embeddings is that past work reports that simpler approaches sometimes work better in the case of software engineering prediction tasks~\cite{hong2022commentfinder}. 
The use of feature-based approaches, and thus low dimensionality representations, has the advantage of low storage complexity and lower computational effort for training.
Text embedding, on the other hand, depending on the approach, can be quite expensive to compute and store, and also increases the training effort.
For the example of BoW, the size of the vocabulary explodes with the size of the corpus increasing. 
LLM encoding approaches usually embed the text into lower dimensions compared to BoW (e.g., around 2k for Bloom). 

While our RQ1 results observed that no specific text-embedding stands out in terms of prediction performance, LLM embeddings report much longer embedding times. In particular, LLM embeddings take .3~seconds per embedding, leading to 1-2 hours of embedding for open-source and 35-48 hours for closed-source datasets. In contrast, BoW embedding the full datasets could be done in 4-12 seconds for open-source and around 5~minutes for the closed-source projects. 
Even if two days of training might not be an issue if a model is kept up for 6 months, we did not perform any data drift analysis to find the optimal retraining frequency, and hence the potential impact of longer embedding time.
Model training on the other hand takes longer with BoW, where the longest training reported for a period was 64~minutes for the BoW embedding and 12~minutes for the LLM embedding, thus 5 times shorter.
We recommend being mindful of the effort needed for each type of encoding, since LLM embeddings require consequently more time overall.

\section{Threats to Validity}
\textbf{Internal validity.}
Review tools can only capture so much, as offline review activities can happen as well. 
In the case of our industrial partner, developers have reported chatting via other tools or doing life reviews in person, which is not captured by their review tool. 
In this study, we thus only focus on what has been recorded in the review tools and aim at predicting those review activities.

\textbf{Construct validity.}
We optimize the results around F1-score, as F1-score is a metric that captures both the recall and precision of the model, which represents best the model's performance with unbalanced datasets. Results could vary slightly if optimized around another metric.

We chose to evaluate the review activity task at file-level, as it is the granularity that is necessary for the use case we considered, file-ordering. Future work could consider the impact of the granularity on the prediction, which is outside the scope of this paper.

We used a zero-shot LLM-encoder as to reduce the computational effort. Fine-tuning the LLMs, though being computationally expensive, could improve the embedding and thus the results.
However, past work has studied LLM fine-tuning for "commented" prediction~\cite{li2022automating} and reported similar results to other Bag-of-Words embeddings~\cite{hong2022should}.  
Larger LLMs might lead to better embeddings, even though the smaller \texttt{bloom-560m} has been reported as more performant than bigger counterparts in some other use-cases~\cite{sharma2023argumentative}, hence it is not uncommon in research to leverage smaller models. As a compromise, we considered \texttt{bloom-560m} as a lower bound of the LLM embedding approach.

We worked on a per-project basis for embedding, training and evaluation, which is the default approach in this domain~\cite{hong2022commentfinder, li2022automating, tufano2022using}. An early bootstrap option, when data is not yet available, might be to initially train on similar datasets (e.g., same coding languages). 

\textbf{External validity.}
The LLM field proposes newly trained/updated LLMs regularly.
We recommend considering recent LLMs for text embeddings to stay on top of the newest findings in case of replication, which could yield better results. The version of Bloom we used is from May 2023. 

Regarding our user study with 29 professionals, the participants only had to review two patches. A large-scale user study over a longer period would be necessary to evaluate the extent to which the hot-spot-based ordering approach can impact the reviews, especially when reviewers start to become aware of the underlying ranking heuristics.


\section{Conclusion}
In this paper, we explored different approaches for embedding code changes to predict review activities. This was motivated by the results of a user study involving 29 professional developers showing how ordering files in patches by prioritizing files that need review activity improves the review process. Indeed, we observe that such ordering approach lead to more review comments (+23\%), that targeted parts needing revisions or comments (+13\% precision and +8\% recall), thus the need for an efficient review automation prediction approach.
While previous work used a Bag-of-Word approach on the added lines and additional features (i.e., \#lines added and \#lines removed)~\cite{hong2022should},
we proposed variations to that embedding. 

First, we improve the textual embedding part by considering embedding the removed lines as a contextual embedding and also considering LLM embeddings. 
Though no specific text embedding stands out in our empirical study on 3 open-source and 2 open-source projects, adding any of our variations improved results compared to the BoW on the added lines in terms of F1-score in 13/15 cases (up to +5\%).
Then, we propose review process feature embeddings, using size- and history-based features.
In practice, this performed at least as good as the textual embeddings in 10/15 cases regarding F1-score, while being less computationally expensive.
Especially closed-source datasets did not show significant difference between text embeddings and the feature-based approaches. 
Finally, when combining the text and feature-based embeddings, our approach reports a median F1-score of 40-62\%, improving the performance significantly with medium to large effect compared to the baseline in all cases (up to +9\% F1-score) with the best combination approach for all tasks. 

Hence, we recommend future work on these types of tasks to consider our text embeddings and review process feature embeddings, as they correlate with significant performance improvements. 
We observed that hot-spot prediction as file-ordering approach reports upto +29\% recall in the first half of patches compared to the default alphanumeric ordering.

\bibliographystyle{ACM-Reference-Format}

\bibliography{acmart.bib}

\end{document}